# Physical probability in the Everett interpretation and Bell inequalities[1]

## Simon Saunders[2]

**Abstract**: I define a notion of local causality *LOC*, closely modelled on Bell's principle, construed as the condition that single-case probabilities cannot be modified by actions at spacelike separation. The new principle, like Bell's, forces Bell inequalities, but with two loopholes: one is retrocausation, known to Bell, but the other is non-uniqueness of remote outcomes, a loophole only for *LOC*, not for Bell's principle.

I also set out a theory of physical probability, applicable to the Everett interpretation, in agreement with the Born rule, and therefore violating Bell inequalities. I show it is consistent with *LOC*. Surprisingly, both loopholes are exploited. I conclude not only that probability in the Everett interpretation involves no action-at-a-distance, but that the observed violations of Bell inequalities is powerful evidence for many worlds.

> The 'many world interpretation' seems to me an extravagant, and above all an extravagantly vague, hypothesis. I could almost dismiss it as silly. And yet...it may have something distinctive to say in connection with the 'Einstein Podolsky Rosen puzzle', and it would be worthwhile, I think, to formulate some precise version of it to see if this is really so. (Bell 1990 p.194)

## 1 Introduction

Alice and Bob are far separated in space. There is a source of spin-systems, produced in pairs, with one from each pair sent to Alice and the other to Bob. They each measure a component of spin in one of two directions, the choice to be made freely at the time of measurement. After many repetitions, they establish contact with Charlie, to whom they send their records of outcomes. Charlie finds those records strangely correlated, in ways that defy local, causal reasoning, but which follow from quantum mechanics.

The setup is due to Bell (1964), as inspired by Einstein, Podolsky and Rosen (1935), following modifications introduced by Bohm (1951); call it the *EPRB setup*. Bell argued that assuming probabilities can be defined for the various outcomes of such experiments, then from local, causal considerations, they should satisfy certain inequalities. The probabilities can be calculated in quantum mechanics and violate those inequalities. The experiments, carried out with increasing precision, confirmed its predictions. Alice, Bob, and Charlie (not their real names) receive the Nobel prize in physics, having shown that Bell inequalities are experimentally violated, and that something is wrong with the local, causal reasoning, that led to them.

The challenge thus posed to Everettian quantum theory has not always been recognised. It will not do to insist that only when Charlie gathers the records, are the inequalities violated, with all the measurements completed in the causal past of a single observer; the point is still to explain their violation. It is not enough to point to Lorentz covariance at the fundamental level, whether of the relativistic quantum state or of local quantum fields; perhaps the issue lies with the emergent reality, and anyway, probability has not yet entered the picture (and the nonlocality is all about probability). The corelative step (for otherwise Lorentz covariance is spoiled), of removing collapse of the state, is insufficient: if the quantum state is something physical, as in Everettian theory, its collapsing by either Alice or Bob involves action-at-a-distance, end of story; but that does not remove the apparent non-locality of probability in the EPRB setup, it removes a non-local way of explaining it. Then what does explain it? To reply that the reasoning involved

---

presupposes a single world, leaves open the question of what reasoning *does* apply, in Everettian terms, to the EPRB experiment.

Partial answers have been given to these questions, including in Timpson and Brown (2002), Bacciagaluppi (2002), Wallace (2012), Tipler (2014), and Brown and Timpson (2016). There is some consensus: like ordinary quantum mechanics, Everettian theory violates *outcome independence*, one of the three key conditions used to derive Bell inequalities (this following Clauser and Horne (1974), Bell (1976)); the condition is implied by what Bell called 'local causality', but appears motivated by Reichenbach's principle of common cause – or at any rate, some doctrine on how correlations may be explained. Brown and Timpson (2016) further argue that the failure of outcome independence in Everettian theory shows only that a Reichenbach-style explanation of correlations fails, with no implication of action-at-a-distance. If it was tacitly built into Bell's definition of local causality, as also suggested by Myrvold et al (2024), that explains the limitations of Bell's principle, and the means whereby there may yet be 'peaceful co-existence' with relativity. Everettian theory is unusual not just by virtue of many worlds, but by virtue of entanglement – and thereby, of non-separability. Bell's principle went wrong in failing to consider a non-separable world, as did Reichenbach's.

The argument is persuasive, but if correct, it shows that many worlds is not really at issue – perhaps another kind of non-separable theory, without many worlds, may similarly violate outcome independence (and so Bell inequalities), with no action-at-a-distance; this chimes with the view that, perhaps, Bell's principle was too strong, as building in an additional demand for explanation of correlations, that may be rejected. Here I suggest an alternative diagnosis: Bell erred in not following through on his guiding intuition, which was the principle of no action-at-a-distance: the principle is about *actions*. Bell instead spoke of invariance under the *specification* of remote events, when it ought to have been remote actions. Doing this, there is no implicit demand for explanation, only of independence from distant actions.

Bell's principle of local causality is thus easily amended, in a way that he might well have found congenial; we do not move outside of his way of thinking.[3] Call the resulting principle $LOC$. Now for my central claim: $LOC$ is consistent with one-world, many-world, and non-separable theories. It rules out none of them. But it forces different conditions in the first two cases – in the many-world case, *conditions insufficient to force Bell inequalities.* The loophole is this: remote experiments need not have unique outcomes. This suggests Everettian quantum theory may well be consistent with $LOC$.

If we are to examine this question seriously, we need an account of probability in Everettian theory, and specifically of objective or (as I shall call it) *physical* probability. Bell was trying to place probability in the causal order, and so are we. This is the other part to my strategy: to define a theory of probability, in this physical sense, applicable to Everettian quantum mechanics.

That this cannot be done, or is anyway unnecessary, is a commonplace in the literature on Everettian theory, including in writings of some of its strongest advocates.[4] Rather than address those claims directly, here I simply present, as it were, a counterexample. It is true that the concept as I define it lacks the attributes of uncertainty, ignorance, and unpredictability, but now the distinction between physical and epistemic probability is important. All those notions are epistemic. It may be that they come in to play *only* with epistemic probability, and rely on features specific to epistemology, like rational agency, the knowing subject, and self-location. Insisting that physical probability must have a comparable feature may well be a kind of category error.[5]

---

[3] Bell often expressed disquiet as to the correct statement of his principle; see Brown and Timpson (2016), Wiseman (2014). I come back to this in the next section.

[4] For example, Deutsch (2016), Brown and Porath (2020), Vaidman (2021). An exception is Wallace (2012), albeit still tied to epistemic probability. I have long argued that probability is physical in Everettian theory; see for example Saunders (2005, 2010).

If this much is granted, if probability in the physical sense does not have to involve uncertainty, then let it at least involve statistics. My proposal is this: in state $\psi$ at time $t$, the probability of projector $P$ is defined, to greater or less precision, by a certain pattern in $\psi$ at $t$, involving a certain choice of basis -- an expansion of $\psi$ in terms of states of equal Hilbert-space norm, call *microstates,* all (or almost all) of which diagonalise $P$. It is a 'microstate-counting rule', in the tradition of Boltzmann.[6] Unlike the decoherence basis, this basis is purely synchronic, defined independent of the dynamics, depending only on $P$ and $\psi$, but like the decoherence basis, permitting of greater or less fine-graining. It is in agreement with the Born rule. Indeed, on embracing a rather more general principle, abstracted from $LOC$, the Born rule can be derived.

In this way I introduce not one, but *two* theories of probability, $\lambda$-MANY and $\lambda$-ONE, where $\lambda$ labels microstates. $\lambda$-MANY is my serious candidate theory of physical probability for Everettian quantum mechanics, as just sketched. The microstates have no independent existence, unlike Everettian branches, which are emergent, dynamically autonomous entities. But in $\lambda$-ONE, as the name suggests, just one of these microstates is produced on each trial, selected at random from an expansion for $P$ in $\lambda$-MANY. This theory also satisfies the Born rule and violates Bell inequalities. It is a one-world, deterministic hidden-variable theory, like pilot-wave theory.

The question arises whether it is so similar to pilot-wave theory as to suffer from the same pathology that first motivated Bell's investigations (as recounted in Bell (1966)): must $\lambda$-ONE, like Bohmian mechanics, involve action-at-a-distance? If so, there is something fishy about $\lambda$-MANY as well, as concealed in the superposition is an underlying nonlocality at the level of the microstates. It would present the same conspiratorial air as pilot-wave theory, which conceals at the level of probabilities (averaging over repeated trials) the action-at-a-distance present in each individual trial.

However, this is not the case. As we shall see, $\lambda$-ONE exploits a *second* loophole in the derivation of Bell inequalities from $LOC$: it involves retrocausality.[7] $LOC$ does not cease to apply in such a case, but the conditions then implied are too weak to force Bell inequalities. $\lambda$-ONE is thus consistent with $LOC$, as is $\lambda$-MANY. Therefore Everettian quantum theory, supplemented by $\lambda$-MANY, is consistent with $LOC$ as well.

This last step is needed if we are to draw any wider conclusions from the observed violations of Bell inequalities. It is one thing to argue from $LOC$ and the experimental evidence to the conclusion that one or both loopholes are exploited in nature; but if $LOC$ rules out Everett, the inference to many worlds is useless – the conclusion that remote experiments do not have unique outcomes is simply baffling. But if consistent with $LOC$, Everettian quantum theory can be used to explain the stunning conclusion: from the observed violations of Bell inequalities, if there is no retrocausality and no action-at-a-distance, remote quantum experiments do not have unique outcomes.

**2 Bell's theorems**

I use the notation of Myrvold et al (2024). Let Alice measure the component of spin of her spin-system in one of the directions $a, a'$ (unit vectors in 3-dimensions) with outcomes $s = \pm 1$, spin-up and spin-down; and likewise let Bob measure the spin of his spin system in one of the two directions $b, b'$, with outcomes $t = \pm 1$. The initial state of the two spin systems is the singlet state, entangling Alice and Bob's experiments. In this $a, a', b, b'$ are all proper names (we suppose the possible choices of directions to be fixed in advance); only $s, t$ are variables.

---

[5] With this joint (physical) probabilities for spacelike separated events are also meaningful, whether or not any localised agent is subsequently able to deliberate upon them. Charlie drops out of the picture.
[6] Boltzmann's microstates were cells of phase space of equal Liouville measure (whereas ours are vectors in Hilbert space of equal norm). Einstein, late in life, singled out this idea of Boltzmann's as of 'outstanding importance' (Einstein 1949 p.43). For its role in the discovery of quantum mechanics, see Saunders (2020); for its relation to Gibbs' concept of probability, see Saunders (2024).
[7] See Friedrich and Evans (2023) for a survey.

Following Bell, we assume there exists a probability distribution $\rho$, determined at the source of the pairs of spin-systems, and probabilities $p_{a,b}(s,t|\rho)$ for Alice and Bob obtaining outcomes $s$ and $t$, having set their parameters to $a, b$ respectively, conditional on $\rho$, with similar functions for parameters $a, b'$, $a', b$, and $a', b'$. Throughout we talk of Bob's experiment as the remote system.

Bell defined his principle as follows (Fig.1):

> A theory will be said to be locally causal if the probabilities attached to values of local beables in a space-time region 1 are unaltered by *specification of values of local beables* in a space-like separated region 2, when what happens in the backward light cone of 1 is already sufficiently specified, for example by a full specification of local beables in a spacetime region 3. (Bell 1990 p.239; emphasis mine)

I propose in its place, changing only the italicised words:

LOC  Probabilities attached to values of local beables in a space-time region 1 are unaltered by *causal changes* in a space-like separated region 2, when what happens in the backward light cone of 1 is already sufficiently specified, for example by a full specification of local beables in a spacetime region 3.

By 'causal change' I mean a dynamically-allowed change. I further cash out this notion in terms of Bob's *free* actions in region 2, dynamical changes consistent with the full specification of region 3. Bell himself spoke in this way and defended this usage against criticism (Bell 1977). He further cited Einstein, quoting, on several occasions, his remark:

> But on one supposition we should, in my opinion, absolutely hold fast: the real factual situation of the system S1 is independent of *what is done* with the system S2, which is spatially separated from the former. (Einstein 1949 p.83, emphasise mine).

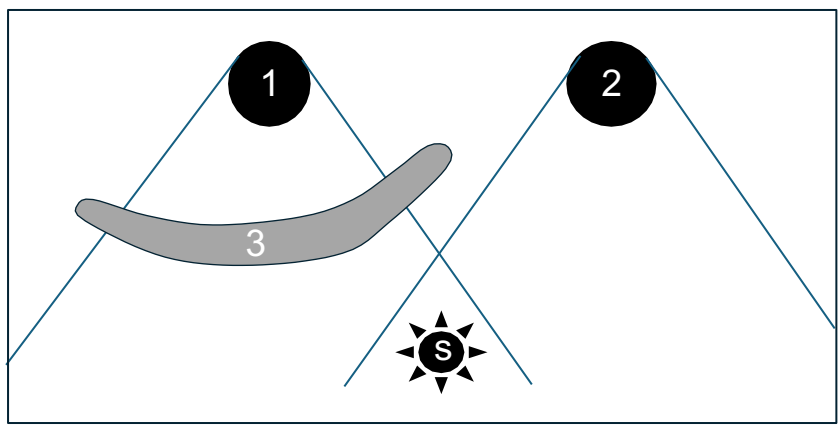

Fig 1: Alice conducts her experiment in spacetime region 1, Bob conducts his in region 2; the source of spin-systems is in region S, in the causal past of Alice and Bob. The region 3 entirely obstructs the past light-cone of 1.

'What is done' denotes action. Bell, in stating his general principle, sought precision at the expense of intuition, and with some foreboding.[8]

The marginal probabilities for Alice and Bob's measurements are

$$p_{a,b}(s|\rho) = \sum_t p_{a,b}(s,t|\rho)$$
$$p_{a,b}(t|\rho) = \sum_s p_{a,b}(s,t|\rho).$$

From $LOC$, Bob's free choice of instrument setting, whether $b$ or $b'$, must leave Alice's probabilities unchanged:

$$p_{a,b}(s|\rho) = p_{a,b'}(s|\rho). \qquad (1)$$

An analogous equation follows for Bob's probabilities, under Alice's change of parameter settings. This condition is parameter independence. Its violation would allow of superluminal signalling: Alice and Bob may have a store of paired spin-systems, and by repeated measurements Alice may test for any change in probabilities due to Bob's remote actions.

We distinguish Bob's action in choosing the spin direction from his action in actually carrying out the experiment. $LOC$ applies to the latter choice as well: if Bob so chooses, his action should leave Alice's probabilities unchanged. *Given that the outcome of his experiment is unique,* then whether it is $t = +1$ or $t = -1$, it is a causal, dynamically allowed change, the consequence of his free action; so $LOC$ then implies:

$$p_{a,b}(s|\rho) = p_{a,b}(s|t,\rho). \qquad (2)$$

Eq.(2) and its cognate is outcome independence. As implied by $LOC$, Eq.(2) forbids a certain kind of action-at-a-distance, and so has nothing to do with explanation - *provided* that Bob's action, in carrying out his experiment, produces a unique outcome. Notice further that on this supposition, the value of $t$ is not controllable, so unlike parameter independence, the condition cannot be tested. And notice finally, that (2) appears forced by Bell's formulation of local causality even if Bob's remote outcome is *not* unique – as it presumably applies on 'specifying' just one of his outcomes.[9]

For future reference, we shall also need the symmetric condition:

$$p_{a,b}(s,t|\rho) = p_{a,b}(s|\rho)p_{a,b}(t|\rho). \qquad (3)$$

For values of $s,t$ for which (3) is non-zero, (3) implies (2), and vice versa, by the definition of conditional probability. We shall call (3) completeness.[10]

The assumption that $\rho$ is independent of the instrument settings of either Alice or Bob is measurement independence, the last condition needed to derive Bell inequalities. It follows automatically on the assumption that $\rho$ is fully controllable – that it is the same, as produced by the state-preparation device, on each trial. We shall come back to this shortly.

Given all this, Bell inequalities, and specifically CHSH inequalities (after Clauser, Horne, Shimony and Holt), follow directly. The relevant conditions can be translated into ordinary quantum mechanics quite easily, with $\rho$ the quantum state, introducing a tensor product structure for the two regions 1 and 2, and computing probabilities using the measurement postulates. Alice and Bob are free to choose local unitary transformations (they control local, time-dependent Hamiltonians, 'free parameters' according to Bell, 'exogenous variables' according to Myrvold et al). On this basis, measurement independence is assured. Parameter independence is easily proved -- this is the quantum no-signalling theorem, that first came to light in Bell's work. But outcome independence is violated when $\rho$ is an entangled state; since a one-world theory, this is action at-a-distance, a change in Alice's probabilities that has resulted from a remote, lawlike change, as produced by Bob's free choice to carry out his experiment. Ordinary quantum mechanics therefore violates $LOC$. All this is exactly as expected.

---

[8] The step, he explicitly advised, should be treated with the 'utmost suspicion', as 'likely to throw out the baby with the bathwater' (Bell 1990 p.239). The baby was Everett.

[9] Brown and Timpson (2026) makes this point: Everettian quantum theory is ruled out by Bell's principle of local causality.

[10] Terminology for the conditions (2), (3) varies wildly. Eq.(3) first appeared in Clauser and Horne (1974), but went unnamed; 'completeness' was introduced in Jarrett (1984). Myrvold et al (2024) calls (3) outcome independence, citing Shimony (1986), and supposes that it is motivated by a no-distant action principle and Reichenbach's principle (the two together implying Bell's local causality principle).

Bell argued the same from his principle of local causality. He gave the example of an atom undergoing alpha particle decay, surrounded by a ring of detectors. If one detector is fired, the probabilities of all the others firing are instantaneously changed. It is similar to Einstein's photon box thought experiment, sketched in correspondence in the 1930s: a photon is in a superposition of two localised states, one state in each box, the two boxes far apart. On opening one of the boxes (and discovering or not discovering the photon) the probability that it will be found in the other box is instantly changed. Both examples involve entanglements (formed as soon as the alpha particle or photon interacts with detectors).[11] Entanglement is all that is needed to violate outcome independence; the violation of Bell inequalities is a different condition. Similarly, when Alice and Bob measure spin components in the same direction, outcome independence is violated, hence Bell's principle (and $LOC$, given a single world), although Bell inequalities are satisfied.

But this raises a puzzling question. Then why is the violation of Bell inequalities important to the question of locality, if action-at-a-distance is more simply illustrated by failures of outcome independence? -- as, arguably, was already obvious to Einstein. One answer is that the latter cannot be directly tested. Another, although still not the whole of it,[12] is that the example of the alpha-particle, and the photon box, and EPRB with parallel parameter settings, can all be explained by hidden variable theories, all perfectly free from action-at-a-distance.

To understand this better, we need the *other* Bell theorem, based on the idea of a deterministic hidden-variable theory, strangely dual to the kind we have been considering. In such a theory, given the hidden variable, denote $\lambda$, the outcomes of the two experiments are uniquely determined. We may use the same notation as before, but now require:

$$p_{a,b}(s,t|\lambda) \in \{0,1\}; \; \lambda \in \Lambda \tag{4}$$

and similarly for $a, b'$, etc., where $\Lambda$ is a suitable measure space, equipped with a probability distribution $\rho$. Nontrivial probabilities of the outcomes are given by averaging over the $\lambda$'s with respect to $\rho$:

$$p_{a,b}(s,t|\rho) = \int_\Lambda p_{a,b}(s,t|\lambda)\rho(\lambda)d\lambda. \tag{5}$$

In this way we recover the quantities taken as basic in our previous treatment, the LHS of (5).

Such a theory is *factorizable* if

$$p_{a,b}(s,t|\lambda) = p_a(s|\lambda)p_b(t|\lambda) \tag{6}$$

---

[11] See the discussion of the Einstein photon-box thought experiment by Vaidman, this volume (in the Fock space representation, the entanglement is obvious). For more on the history, see Bacciagaluppi and Crull (2023). Brown and Timpson (2016) denies that this example involves entanglement.

[12] A local deterministic hidden variable theory, if it makes sense of probabilities other than 0 and 1 in the single case, must still have them change at a distance in these examples. Why is this acceptable? I suggest because they are then read as epistemic probabilities, and as such not bound to the causal order in the way of Bell's principle of local causality, or my variant $LOC$.

From this and (5), for certain choices of parameter settings $a, a', b, b',$ and assuming measurement independence, Bell first derived an example of his famous inequalities.

We can connect factorizability to our previous conditions quite simply. Concerning $LOC$, we may regard a deterministic hidden variable theory as the limiting case in which all probabilities are 0s and 1s. Outcome independence conditional on $\lambda$ follows automatically (because $\lambda$ determines the outcomes uniquely); parameter independence, conditional on $\lambda$, in this a one-world theory, follows from $LOC$:

$$p_{a,b}(s|\lambda) = p_{a,b'}(s|\lambda) \tag{7}$$

Factorizability follows, hence Bell inequalities. Bell (1964) argued directly for (7), quoting Einstein (the quotation already given). But now measurement independence takes a stronger form, in that the $\lambda$ produced by the state-preparation device is not controllable at the time of emission, unlike $\rho$, so conceivably could depend on Alice and Bob's later choices of parameter settings – or conversely, could dictate those settings. The latter Bell called *superdeterminism*: Bob is not free to choose his instrument settings -- his choice is dictated by $\lambda$. Such a theory may be factorizable yet (5) fails, whereupon Bell inequalities cannot be derived – this the loophole granted by Bell. The other way of putting it is in terms of retrocausation: Bob freely chooses his settings, but his choice dictates the $\lambda$ emitted at the source. They are two sides to the same coin, it is the same loophole.

Any deterministic hidden-variable theory that reproduces quantum probabilities and satisfies measurement independence must violate (7). This answered the question Bell had previously posed in his (1966): Bell knew that quantum probabilities could be reproduced by a deterministic hidden-variable theory, namely pilot-wave theory (he cited Bohm's 1952 papers), but that explicitly involved action-at-a-distance; might there be some other theory of this type, that is free of it? His answer in Bell (1964)[13] was negative; any such must violate (7), just like pilot-wave theory.

The previous theorem, for single-case probabilities other than 0 and 1, applied to what in Bell's time was called an 'indeterministic' hidden variable theory (this included quantum mechanics, with the hidden variable $\rho$ replaced by the quantum state). This is the strange dualism previously alluded to: outcome independence is violated (satisfied), and parameter independence is satisfied (violated), in indeterministic (deterministic) hidden variable theories, like ordinary quantum mechanics (Bohmian mechanics).

Many have called attention to the special status of the two Bell theorems.[14] We shall see consider them again shortly, with $\lambda$-MANY replacing ordinary quantum theory, and $\lambda$-ONE replacing pilot-wave theory.

### 3 The theory $\lambda$-MANY[15]

We assume the basic framework of Everettian quantum mechanics: there exists a unique vector state $\psi \in \mathcal{H}$, 'the universal wave-function', the superposition principle has unlimited scope, and the dynamics is always unitary. No more detail is needed. $\psi$ is not assumed to be normalised, and may have any amplitude and phase.

---

[13] It seems there was no exciting retrocausation: Bell's earlier submission was lost by the editorial office.
[14] See in particular Wiseman (2014). Incidentally, this lists among axioms needed to derive Bell inequalities, alongside causal ones, the axiom that 'exactly one outcome happens to Alice and Bob, on each trial'. If this axiom is really needed, and the other axioms are consistent with Everettian theory, it adds support to the conclusion of §6.
[15] Here I closely follow Saunders (2025); there are several differences from my (2024), notably, there is no constraint that microstates are fine-grainings of the decoherence basis.

We consider expansions of $\psi$ into microstates, orthogonal vector states of equal Hilbert-space norm (so equal amplitude - 'amplitude' means norm). For any such expansion, denote by $\Lambda_\psi^n$ the set of $n$ microstates involved ($n$ must be finite since $\psi$ must have finite norm). We define probabilities for projectors $P$ on $\mathcal{H}$ as follows.

To begin with, suppose there is an equiamplitude expansion in microstates that are all eigenstates of $P$. That is, suppose:

$$\psi = \xi_1 + \ldots + \xi_n \; ; \quad \|\xi_j\| = \|\xi_k\|; \quad P\xi_j = \xi_j \text{ or } 0; \quad j, k = 1, \ldots, n.$$

In such a case, we shall say $\Lambda_\psi^n$ is *adapted* to $P$. Suppose $m$ microstates lie in the range of $P$, so $n - m$ lie in $I - P$; we then say that the probability of $P$ is $m/n$; its frequency in $\Lambda_\psi^n$ is $m/n$. It trivially follows that $\Lambda_\psi^n$ is adapted to each of the projectors $P_{\xi_k}, k = 1, \ldots, n$, and that each has probability $1/n$. Implicit, then, in this picture of probability, is the condition that microstates be equiprobable.

We can rephrase the rule in terms of states. Let the superposition of $m$ microstates equal $P\psi$, and of $n - m$ equal $(I - P)\psi$: if the microstates have equal probability, then the probability of $P\psi$ is $m/n$. Similarly, the probability of $P_{\xi_k}\psi = \xi_k$ is $1/n$. (In this form, it applies to projectors of a slightly more general nature – when, for example, individual microstates are not eigenstates, but there is a partitioning of $\Lambda_\psi^n$, with the sum of vectors in each partition an eigenstate of $P$; but this need not concern us here.)

The Born rule for $P$ in the state $\psi$ agrees with our microstate-counting rule:

$$\frac{\|P\psi\|^2}{\|\psi\|^2} = \frac{\|P(\xi_1 + \ldots + \xi_n)\|^2}{\|\xi_1 + \ldots + \xi_n\|^2} = \frac{\|(\xi_1 + \ldots + \xi_m)\|^2}{\|\xi_1 + \ldots + \xi_n\|^2} = \frac{m}{n}.$$

The same equation follows for any other equiamplitude expansion $\Lambda_\psi^{n'}$ of $\psi$ into $n'$ microstates adapted to $P$, yielding probability $m'/n'$. Therefore $m'/n' = m/n$. For projectors like these, $\lambda$-MANY is simple.

What of expansions that are not adapted to $P$? They yet place *bounds* on the probability of $P$ – an upper-bound, defined by the number of microstates that have eigenvalue 0, a lower-bound, defined by the number of microstates that have eigenvalue 1. There is a grey zone in between, consisting of microstates that are not eigenstates of $P$ - call them 'Schrödinger-cat states for $P$'. The larger this interval, the less informative the expansion, but since all the microstates are equiprobable, these bounds on probabilities can never contradict each other. This can be proved directly: for any expansion, the interval between upper and lower bound for the probability of $P$ must always contain the Born-rule quantity, so they can never be disjoint.[16] Call such pairs of bounds 'imprecise probabilities'.[17]

This picture of probability is clearly akin to frequentism, indeed finite frequentism, in (classical, one-world) philosophy of probability, where likewise the fraction of an appropriate finite ensemble, that has property $P$, is the probability of $P$ in that ensemble. Likewise elements of the ensemble are implicitly taken as equiprobable. And likewise, there may be borderline cases, so imprecise probabilities may enter as well. The difference lies in the nature of the ensemble: classically, it is spread out across space and time, involving many repeated measurements; quantumly, it is spread out across a superposition at a single place and time, involving a single measurement.

With this difference, the frequentist picture of probability in terms of the balls-in-urn model can be taken over as well. Let the balls be marked either blue or green. Classically, the fraction of balls in the urn that are green gives the probability that a ball drawn randomly from the urn is green. Quantumly, all the balls are drawn from the urn, and a measurement of colour partitions them into those that are green, and those that are blue, with probability of green given by the fraction that are green, and similarly for blue. The partitioning is done by a quantum measurement apparatus under the unitary dynamics (and is carefully engineered to that end). One or more of the balls may be ambiguous as to colour – we have a Schrödinger cat; there may be no measurement possible whose outcomes correspond to individual balls (the differences between the balls being too tiny to discriminate) -- illustrating that microstates are not Everettian worlds, or branches, which (for the sake of argument) we suppose must differ macroscopically. Worlds have different probabilities according to the proportion of microstates that superpose to give those worlds, because the microstates are equally probable.[18] Epistemic probability comes into the picture if we imagine there are observers, witnessing the draw, finding themselves partitioned as well, some in the green partition, some in the blue; at least momentarily, they will be uncertain as to which partition they are in (they will have 'Vaidman uncertainty', Vaidman (1998)).

---

[16] The proof is easy; see Saunders (2024) (and for more background on imprecise, or what I there called 'interval' probabilities).

[17] Following e.g. Bradley (2019). This literature is almost exclusively concerned with epistemic probability (and naturally, restricted to a single world).

More is needed to obtain probabilities as real and not just rational numbers. To that end, suppose $\dim \mathcal{H} = \infty$, and consider projectors $P$ on $\mathcal{H}$ with $\dim P = \dim(I - P) = \infty$, that are otherwise arbitrary, defined without any reference to $\psi$. Projectors like these are needed to describe any system with a continuous degree of freedom, like an atom in space; a projector $P$ onto the closure $\Delta$ of any open set in space, no matter how small, has infinite dimension (the Hilbert space $L^2(\Delta, dx)$ is infinite-dimensional), and so does $I - P$. Arguably, *any* realistic physical system falls into this category, as involving at least one continuous degree of freedom.[19] (This need for projectors of infinite dimension is often concealed. If we separate out spin-degrees of freedom in terms of a tensor product structure, we have projectors of the form $P \otimes I$, where $P$ is of finite and perhaps very small dimension; if we ignore the centre-of-mass degrees of freedom, spin-systems can be treated entirely in the framework of finite-dimensional Hilbert space. But spin-systems have spatial properties -- or at least, the ones in the EPRB setup do -- so the underlying projectors are really of the form $P \otimes I$, as likewise their complements $(I - P) \otimes I$, which are always infinite-dimensional.)

I state without proof two key properties of equiamplitude expansions in Hilbert space:

EE1  If $\dim \mathcal{H} \geq 3$, for any vector $\psi \in \mathcal{H}$ and for any $n$, $2 \leq n \leq \dim \mathcal{H}$, there exists a continuous infinity of equiamplitude expansions of $\psi$ in $n$ microstates

EE2  If $\dim \mathcal{H} = \dim P = \dim(I - P) = \infty$, then for any state $\psi \in \mathcal{H}$ and any $n \geq 2$, there exists an equiamplitude expansion of $\psi$ into $n$ microstates, at most one of which is not an eigenstate of $P$.

The first can be proved by simple induction (Saunders 2025), the second by explicit construction. We shall still say finite expansions of the form EE2 for large $n$ are adapted to $P$, ignoring the Schrödinger-cat state, as $n$ can be taken as arbitrarily large, and in the limit $n \to \infty$, the probability thus defined is precise. In this way we obtain a vastly expanded class of projectors for which precise probabilities can be defined – as real numbers.

This use of the infinite limit is common to all forms of frequentism that aspire to define probabilities as real and not just rational numbers, but there are two important differences from the classical case. Classically, $n$ is the number of trials of an experiment; of course, the $n \to \infty$ limit is unphysical, but even moderately large numbers are unphysical – a billion tosses of a single coin is unphysical. The $n \to N_A$ limit, where $N_A$ is Avogardro's number, is definitely unphysical, for any kind of classical trial. But in the quantum case it poses no difficulty; nor does the limit

$$n \to N_A^{(N_A^{N_A})}$$

and so on, without bound, given that $\dim \mathcal{H} = \infty$. The superposition that results, no matter of how many microstates, is always a representation of the same physical state $\psi$ at the same instant of time, for the same single experiment.

---

[18] The suggestion that worlds be considered equiprobable, made in Graham (1973), is of course disastrous, but fortunately has no good argument. See Saunders (2024), for a more plausible variant on Graham's branch-counting rule, which also fails, refuting Khawaja (2023).

[19] This point is open to challenge, particularly in quantum gravity; but it seems appropriate that the physics of the continuum, for physical probability as for space, should stand or fall together. Still, $\lambda$-MANY applies, but 'most' projectors would have only imprecise probabilities (for a Hilbert space with dimension of order the Beckenstein bound, of course, the imprecision will be vanishingly small).

The second and even more important difference is that classically, the frequency of $P$ in an infinite ensemble, as defined by taking the limit $n \to \infty$,[20] may differ arbitrarily from its frequency in any finite sub-ensemble. If the subensemble is chosen, somehow, at random, the frequency of $P$ in the sub-ensemble will *probably* equal its limit frequency -- this the law of large numbers -- but probability will then depend on this notion of randomness. In the quantum case, none of this applies. The limiting frequency of $P$ necessarily satisfies the bounds placed on it by any expansion in microstates $\Lambda_\psi^n$, for any $n$ – bounds which, by EE2, for expansions adapted to $P$, have imprecision of at most $1/n$. Those bounds are not *probably* correct, they are correct simpliciter. And no *random* selection of sub-ensemble is involved, in the choice of an equiamplitude expansion; it is fine-tuned to $P$.

Consider now the scope of $\lambda$-MANY. The limit of the sequence of imprecise probabilities for $P$, as secured by EE2, is a real number, equal to the Born rule quantity for $P$ in the state $\psi$. This includes the Born rule in ordinary quantum mechanics as applied to any macroscopic experiment; since apparatuses have spatial degrees of freedom, their properties must satisfy EE2. Therefore $\lambda$-MANY includes the Born rule as ordinarily applied. It further extends to projective probabilities as defined in decoherence theory – projectors 'diagonal in the decoherence basis' (expansions of $\psi$ adapted to such projectors will likewise be diagonal in this basis) – and in this way includes more sophisticated uses of the Born rule, including decoherence-based Everettian quantum theory.[21] But $\lambda$-MANY goes well beyond this too, for it applies equally to isolated microscopic systems – to any quantum system with at least one continuous degree of freedom – whether or not there is decoherence. The probabilities thus defined are always instantaneous probabilities. It is the question of how they align over time that brings in the dynamics and ultimately returns us to decoherence theory.

All of this is promissory, however, for integral to $\lambda$-MANY is the equality of probability of states of equal Hilbert space norm. However natural, it is in effect an *assumption*.

Can we not appeal to the Born rule, legitimated, perhaps, through the Deutsch-Wallace decision-theory approach? No – for neither has the required scope. In any case, we are attempting to hold physical and epistemic notions of probability strictly apart. Might we simply posit the Born rule, to hold with complete generality? But that goes well beyond $\lambda$-MANY, and would reduce it, at best, to an explanatory picture.

I propose instead the principle:

INVARIANCE    The physical probability of X cannot be changed without a causal change in X.

On translating into a physical theory, it is clearly defeasible. It can moreover be read as an abstraction from $LOC$, essentially contracting region 3 to the boundary of region 2 in Fig1, and in this way be translated into the spacetime framework of the EPRB experiment. As such it gives the same results as *LOC*.[22]

---

[20] Called 'hypothetical frequentism', in the terminology of Hájek (1996) – but as we have just seen, classically, even modestly-large finite ensembles are hypothetical as well.

[21] There is obviously more to say about POV measures, and expectation values of self-adjoint operators more generally.

Translated into Everettian quantum theory, we take X to be a vector state, one of several in a superposition, each of which is assigned a physical probability; and we further take causal change to mean unitary change. Thus, let $\varphi$ occur in some expansion $\Sigma$ of $\psi$, with physical probability $\mu_{\psi_\Sigma}[\varphi]$; INVARIANCE requires that if $\varphi$ is unchanged by a unitary transformation $U$, then its probability is unchanged:

$$\varphi \in \Sigma;\ U\varphi = \varphi \Rightarrow \mu_{U\psi_{U\Sigma}}[\varphi] = \mu_{\psi_\Sigma}[\varphi].$$

From this it is easy to prove the theorem (Short 2023, Saunders 2025):

$$\varphi, \eta \in \Sigma, \qquad \|\varphi\| = \|\eta\| \Rightarrow \mu_{\psi_\Sigma}[\varphi] = \mu_{\psi_\Sigma}[\eta].$$

But then it follows, when $\Sigma$ is an equiamplitude expansion of $\psi$, that the states in the expansion (microstates) must have equal physical probability – the needed justification for $\lambda$-MANY. Given that $\lambda$-MANY satisfies INVARIANCE quite generally, with X given by $P\psi$ for any $P$, we have a proof of the Born rule in our extended sense.

## 4 $\lambda$-MANY and parameter independence

$\lambda$-MANY applies to the EPRB experiment similarly as does ordinary quantum mechanics, with $\rho$ given by the quantum state $\psi$, replacing the measurement postulates by microstate counting, with no collapse of the state. The condition that Alice and Bob are causally isolated is made out as usual, by use of a tensor-product $\mathcal{H} = \mathcal{H}^A \otimes \mathcal{H}^B$, with Hilbert space $\mathcal{H}^A$ for Alice and her apparatus and spin system, $\mathcal{H}^B$ for Bob and his apparatus and spin system, and by the condition that Alice and Bob freely control their local unitaries, of the form $U \otimes I$ for Alice, and $I \otimes U$ for Bob.

Measurement outcomes (certain configurations of their apparatuses) are represented by projectors, whereby $P_s^a$ represents Alice's measurement of spin in direction $a$ with outcome $s$, where $s \in \{+1, -1\}$, and similarly $P_t^b$ for Bob's measurement in direction $b$ with outcome $t \in \{+1, -1\}$. Thus $P_s^a \otimes P_t^b$ represents Alice's measurement outcome $s$ and Bob's measurement outcome $t$ on measuring spin in directions $a, b$ respectively.

As projectors for macroscopic pointer positions, they satisfy EE2. The fraction of microstates in $P_s^a \otimes P_t^b$ in state $\psi$ is a perfectly objective feature of reality, albeit involving spacelike events. it does not need Charlie to validate it, or to turn it into a physically meaning probability, as it has nothing to do with epistemology.[23] Of course, it is important that the data Charlie eventually collects from Alice and Bob *will* corroborate these joint probabilities -- and so it will, in the sense of the law of large numbers, using $\lambda$-MANY. Questions of which of Alice's descendants (after multiple experiments) can eventually meet up with which of Bob's may be fascinating, but we know the answers must be consistent with the Born rule.

Parameter independence is obviously satisfied. Let Alice measure spin in direction $a$, and consider outcome s, represented by the projector $P_s^a$ on $\mathcal{H}^A$, and by the projector $P_s^a \otimes I$ on $\mathcal{H}$, both of infinite dimension. Any equiamplitude expansion adapted to $P_s^a \otimes I$ will do to determine its probability – and is clearly independent of Bob's instrument setting. A fortiori, no change in Bob's settings can change this expansion, or the fraction of microstates in $P_s^a \otimes I$.

---

[22] This needs to be shown; a start is made in Saunders (2025), which also contains the proof that $\lambda$-MANY satisfies INVARIANCE in full generality.

[23] According to Brown and Timpson (2016 p.112), 'it is only in [Charlie's] region that measurement outcomes for the individual measurements on either side can become definite with respect to one another'; prior to that the Born rule quantities (they appeal to the Born rule to define probabilities) are 'only formal statements, regarding what one would expect to see, were one to compare the results of measurements on the two sides'. In this epistemic and physical notions of probability are intertwined.

The point is equally obvious from the Born rule. However, caution is needed, *since it is obvious in pilot-wave theory as well.* Pilot wave theory satisfies parameter independence conditional on $\rho$, Eq.(3), but not conditional on $\lambda$, Eq.(7) -- the latter showing the underlying action-at-a-distance in the theory. It is masked by averaging over $\rho$ in Eq.(5), contributing to its air of conspiracy.[24] But now it seems the same must be true of the individual microstates in $\lambda$-MANY, in expansions adapted to Alice and Bob's experiments. For those microstates also assign measurement outcomes deterministically, so satisfy outcome independence, conditional on the individual microstate; if they satisfied parameter independence, in the sense of Eq.(7), factorizability would follow, and on averaging, Bell inequalities would be forced. It seems individual microstates in $\lambda$-MANY must fail parameter independence, just as does pilot-wave theory.

A similar question arises for ontology in Everettian quantum theory; worlds are nonlocal in some sense, but the superposition of worlds is local; Everett's relative states are nonlocal in some sense, but their superposition is local. Does the same apply to microstates in $\lambda$-MANY, and is the kind of nonlocality, action-at-a-distance?

To answer this question, let us define a new one-world hidden variable theory, call $\lambda$-ONE. We suppose we have an expansion of $\psi$ adapted to $P$ (for the EPRB set-up, this will be of the form $P_s^a \otimes P_t^b$), and thus a set of microstates $\Lambda_\psi^n$ for some large $n$. The new theory then says that preparing the state $\psi$, in fact just one of the microstates $\lambda \in \Lambda_\psi^n$ is produced at the source, chosen at random (meaning with uniform probability distribution $\rho = 1/n$ on $\Lambda_\psi^n$). This microstate assigns $P$ either 0 or 1 (we take $n$ as sufficiently large so that the chance of obtaining the Schrödinger-cat state for P is negligible). Since constancy of $\rho$ on $\Lambda_\psi^n$ is the *correct* probability distribution, according to $\lambda$-MANY, the probability of $P$ in $\lambda$-ONE agrees with its probability in $\lambda$-MANY, and hence with the Born rule.[25]

By the reasoning just sketched, $\lambda$-ONE, like pilot-wave theory, must fail parameter independence in the sense of (6). And indeed, parameter independence conditional on $\lambda$ clearly does *not* hold – or rather, it simply fails to apply. The $\lambda$ on the two sides of Eq.(7) cannot be identical. On the LHS, it is a microstate from an expansion adapted to $P_s^a \otimes P_t^b$; on the RHS, it is a microstate from an expansion adapted to $P_s^a \otimes P_t^{b'}$. Since $P_t^b$ and $P_t^{b'}$ differ macroscopically, they are disjoint and have no eigenstates in common (here and in the sequel, we ignore the Schrödinger-cat state).

To spell this out, the two sets of spin microstates for Bob's two choices of parameter settings, are of the form

$$\Lambda_\psi^{n,a,b} = \{\phi_s^{a,k} \otimes \chi_t^{b,k}; \ k = 1, \ldots, n; \ s, t = +1, -1 \} \tag{8}$$

$$\Lambda_\psi^{n,a,b'} = \{\phi_s^{a,k} \otimes \chi_t^{b',k}; \ k = 1, \ldots, n; \ s, t = +1, -1\} \tag{8'}$$

where $P_s^a \phi_s^{a,k} = \phi_s^{a,k}$, $P_{+1}^a \phi_{-1}^{a,k} = 0$, etc. They have no microstates in common. $\lambda$-ONE is *ontologically* contextual;[26] the hidden-variable $\lambda$ itself depends on the context, and if Bob changes his parameter settings, he must change $\lambda$ as well.

May he do so, consistent with *LOC*? Observe that these microstates amount to a pair of hidden-variables, of the form $\phi_s^a$ for Alice and $\chi_t^b$ for Bob, in a fully separable description, and that there is a 1:1 correspondence between the microstates for Alice's outcomes, in the two ensembles; in particular, the fraction of microstates with $s = +1$ is the same, in (8) and (8'), as likewise with $s = -1$ (we know this is so from $\lambda$-MANY). It follows that Bob, in mysteriously switching between

---

[24] But not if developed as a theory in which 'quantum non-equilibrium' is possible, as in Valentini (2025).

[25] But it cannot be justified by INVARIANCE. The proofs of Short (2023) and Saunders (2025) break down if just one among the states in a superposition is distinguished, as in $\lambda$-ONE.

[26] Not just contextual in its value-assignments; as a simple corollary to Gleason's theorem, there is no non-contextual mapping of projectors to 0s and 1s. Bell (1966) discussed this in detail; so did Shimony (1984).

(8) and (8') (which we suppose occurs with his choice of instrument setting), does not alter Alice's probabilities, nor need it change her microstate in region 3. It follows that $\lambda$-ONE is consistent with $LOC$ – it exploits the loophole of retrocausality. Parameter independence conditional on $\lambda$ is the wrong condition, when $\lambda$ can be changed retrocausally; but when $\lambda$ has a bipartite structure, as in $\lambda$-ONE, it can be replaced by a different condition, that follows from $LOC$, and which $\lambda$-ONE satisfies, in the manner just shown.

Our serious theory of probability is $\lambda$-MANY. From that perspective the main lesson to be learned from $\lambda$-ONE is that there is no underlying action-at-a-distance, at the level of individual microstates, concealed on taking the superposition. All well and good. But is not the retrocausality concealed instead, and does this not also hint at some kind of conspiracy?

No. In $\lambda$-MANY, the microstates used in defining the probabilities of Alice and Bob's outcomes, appropriate to their choice of settings, do not have to be retrodicted to the source, unlike in $\lambda$-ONE. There is no reason to do so.[27] The causal story is the usual one: $\psi$ unitarily evolves from the source to the moment of Bob's measurement, where, depending on his choice of instrument settings, and given the careful engineering of his apparatus, and given that he actually chooses to carry out his experiment, branching occurs, and a superposition of measurement outcomes results. This is the only dynamics that is operating: the Schrödinger equation evolving forward in time. The point of representing the superposition of outcomes that results, after the measurement, in an expansion of microstates adapted to those outcomes, is to define their probabilities, not their causal provenance. In $\lambda$-ONE, the two cannot be separated.

## 5 $\lambda$-MANY and outcome independence

The key question for $\lambda$-MANY is outcome independence, Eq.(2). Consider first the equivalent condition of completeness, Eq.(3). Suppose that $\psi$ is a product state of the form $\psi = \phi \otimes \chi$. Consider the probability of $P_s^a \otimes P_t^b$ in $\psi$. We expand $\phi$ and $\chi$ in equiamplitude microstates that diagonalise $P_s^a$ and $P_t^b$ respectively, as before; but now we keep track of their numbers, so let there be $n_a$ and $n_b$ respectively. Of these, let $m_a$ have eigenvalue +1 for $P_s^a$, and let $m_b$ have eigenvalue +1 for $P_t^b$, with all the rest eigenvalue 0. That is, on choosing a convenient ordering, we consider the two expansions (now labelling vectors by eigenvalues 1,0 of $P_s^a$ and $P_t^b$ rather than values of spin):

$$\phi = \phi_{+1}^1 + \cdots + \phi_{+1}^{m_a} + \phi_0^{m_a+1} + \cdots + \phi_0^{n_a} \tag{9}$$

$$\chi = \chi_{+1}^1 + \cdots + \chi_{+1}^{m_b} + \chi_0^{m_b+1} + \cdots + \chi_0^{n_b}. \tag{9'}$$

The product $\phi \otimes \chi$ then consists of $n_a \cdot n_b$ equiamplitude microstates, each an eigenstate of $P_s^a \otimes P_t^b$, the $j \cdot k^{\text{th}}$ of which is one or other of:

$$\phi_{+1}^j \otimes \chi_{+1}^k; \ \phi_{+1}^j \otimes \chi_0^k; \ \phi_0^j \otimes \chi_{+1}^k; \ \phi_0^j \otimes \chi_0^k \tag{10}$$

Of these $m_a \cdot m_b$ have eigenvalue +1, the rest 0, so $P_s^a \otimes P_t^b$ has probability

$$p_{a,b}(s,t|\psi) = \frac{m_a \cdot m_b}{n_a \cdot n_b}. \tag{11}$$

Similarly, the probabilities of $P_s^a \otimes I$ and $I \otimes P_t^b$ are

$$p_{a,b}(s|\psi) = \frac{m_a}{n_a}, \ p_{a,b}(t|\psi) = \frac{m_b}{n_b}.$$

Their product is Eq.(11), so completeness, Eq.(4), is satisfied, hence too outcome independence Eq.(3): the probability of Alice's outcome spin-up, conditional on Bob's outcome spin-down, is the same as that conditional on Bob's outcome spin-up -- and is the same if Bob performs no measurement at all. But this reasoning has nothing to do with explaining correlations between Alice and Bob – on the contrary, there *are* no correlations, when completeness is satisfied.

---

[27] It may yet be motivated as a way of introducing pre-measurement uncertainty, as in Saunders (2010, 2021); but that, as I now see it, concerns only epistemic probability.

This same reasoning shows when, precisely, completeness is violated: when $\psi$ is *not* a product state. It will still have expansions in terms of product states of the form (10), adapted to $P_s^a \otimes P_t^b$, but it will no longer be possible to write these expansions as the product of two expansions (9), (9') – which is what completeness requires. The microstates may be paired with one another in a more interesting way – for example, half with Alice spin-up and Bob spin-down, and half with Alice spin-down and Bob spin-up. This is the case when Alice and Bob measure spin in the same direction $a = b$, for the singlet state of spin.

That pairings like this should exist, all in a superposition, that cannot be written as the product of two sets of states, each in a superposition, is fundamental to quantum mechanics, and the way that correlations there arise. It is entanglement. Whether it calls for explanation presumably depends on the pairings.[28] In the EPRB setup there *is* a common cause: the entanglement is prepared by a device in the causal past of Alice and Bob, and it is precision-engineered to build in these special pairings, so there is an explanation. There is no distant action, but there is distant correlation-- correlations that are preserved over large distances to be sure, but correlations by design. It is just a mistake to think that if $\psi$ explains some correlation between Alice and Bob, then completeness must be satisfied. To take the photon-box experiment, the initial entangled state explains why Alice finding the photon is correlated with Bob not finding it, and vice versa (for it is the superposition of the two pairings), but completeness is violated.

Completeness is equivalent to outcome independence; when $\psi$ is entangled, $\lambda$-MANY violates the latter as well. This too is clearly related to co-existence in a superposition. If there is a unique outcome to Bob's experiment, then it is a causal, law-like change, brought about by his free choice in conducting the experiment, no matter that he does not control the outcome. $LOC$ says that causal change cannot change Alice's probabilities -- this is outcome independence. But if Bob's experiment causes *both* outcomes, the verdict is just as clear: Alice's probabilities should be unchanged conditional on both outcomes obtaining, in a superposition, not just on one of them. To produce just one, in Everettian theory, is impossible; $LOC$ does not require that Alice's probabilities be unchanged, if Bob performs a miracle.

It is a moot point as to how the new condition is to be framed, but the safe and easy candidate is to add to the schema a label for the 'start' button of each experiment (assuredly there is one), and conditionalize on that, for it produces a deterministic change. But that only augments parameter independence; no Bell inequalities can be derived in that way. Of course, translated into $\lambda$-MANY, we do know how to describe the multiplicity of outcomes, and we can calculate Alice's probabilities conditional on the production of that remote superposition. Rather obviously, they are unchanged, and the same as if Bob pushes no button at all.

Of course, we *can* conditionalize Alice's probability for outcome *s* on just one of Bob's outcomes, say spin-up; but that is Alice's probability for *s* in Bob's spin-up world, the fraction of microstates with outcome *s* for Alice's measurement among all microstates with spin-up outcome for Bob. There is no reason why that fraction should be the same as the fraction in Bob's spin-down world, computed with respect to an entirely disjoint ensemble, microstates with the spin-down outcome for Bob. No more is it the same as that computed among all microstates, the unconditional probability for Alice's outcome *s*. If Bob's experiment produces only a single outcome, there is only a single ensemble, and the three probabilities must be the same - and change in such a way as to violate $LOC$.

---

[28] I am unpersuaded that it needs to be explained in general – or eliminated, as in the search for a separable formalism for quantum mechanics (especially not if it complicates or undermines a state-based ontology). Yet there is clearly something there; see the chapter by Bédard, this volume.

## 6 What Bell did

There are several lessons to be drawn from the observed violations of Bell inequalities. One concerns the *way* they are violated, as the parameters are varied; when the inequalities are saturated, the change in correlation functions as the angle $\theta$ is changed (leading to violations) is quadratic in $\theta$. If the probability space were a simplex, it would necessarily be linear in $\theta$. Bell pointed this out (1987 p.81-86); see also Pitowsky (1989). The Block sphere, in quantum mechanics, is not a simplex; neither, on variation of $a$, is $\Lambda_\psi^{n;a}$.

Once we assume quantum mechanics, the significance of Bell-inequality violations is rather different. It is a marker of entanglement, since any product state satisfies completeness, as we have just seen. The recent interest in inequality-violating probabilities as measured in high-energy physics experiments is mainly directed to tests for entanglement. However, there are other signatures of entanglement in these regimes that may be more easily detected, in particular the Peres-Horodecki criterion; see e.g. Barr et al (2025).

Violations of Bell inequalities have also been obtained for systems and experimenters separated by large distances (across Lake Geneva, for example). That is noteworthy, but of course evidence that entanglements are preserved over enormous distances is independently available – in interference effects involving gravitational lensing, for example – rather dwarfing this example.

The greater importance of the evidence of violations of Bell inequalities is the generality of the arguments used to derive them. In §2 we concluded

$$(LOC \land \neg RET \land UNIQUE) \to BELL \quad (12)$$

where $\neg RET$ is the banal assumption that causality is always in the forward direction in time, and $UNIQUE$ is the equally banal assumption that remote experiments have unique outcomes. Both conditions are part of our pre-theoretic understanding, independent of quantum theory altogether. It is true that the negation of $UNIQUE$ is *not* of this form – we hardly know what it *means* for uniqueness to fail, without a more detailed theory – but it is not obviously self-undermining (the non-uniqueness, after all, concerns *remote* experiments). Given which, and given that the violation of Bell's theorem is an empirical fact, (12) may be better cast as the logically equivalent sentence: [29]

$$(LOC \land \neg RET \land \neg BELL) \to \neg UNIQUE. \quad (13)$$

Were Everettian quantum theory itself ruled out by $LOC$, the inference would be, at best, suggestive ('search for a new theory in which remote outcomes are not unique, consistent with $LOC$'). But since as we have seen $\lambda$-MANY is consistent with $LOC$, the inference is a new and compelling argument for Everett.

To argue instead:

$$(LOC \land UNIQUE \land \neg BELL) \to RET \quad (14)$$

suffers from the difficulty that we lack the analogue of Everettian quantum theory, explaining (13); what is the retrocausal theory, explaining (14)? -- we do not know what retrocausation involves more generally. There are suggestive examples, to be sure, but like $\lambda$-ONE, they are highly stylised. (14) suggests, rather, a new research programme.

---

[29] Lev Vaidman has long argued that from quantum experiments, no action-at-a-distance, and no retrocausation, many-worlds follows; see his chapter, this volume. However, his argument (here and elsewhere with similar effect) relies on the (deterministic) GHZ experiment, not the EPRB experiment, and says nothing about physical probabilities other than 0 and 1 (indeed there are no such things, according to Vaidman). Waegell and McQueen (2020) make a similar argument, also on the GHZ state, rather than Bell; for a reply, see Faglia (2024).

Those insisting on *UNIQUE* and ¬*RET* must give up *LOC*, the only remaining possibility, unless they wish to call in question the judgment of the Nobel prize committee. They will conclude

$$(\neg BELL \wedge \neg RET \wedge UNIQUE) \rightarrow \neg LOC.$$

It is on better ground than (14), in that there is at least one worked-out quantum theory that involves action-at-a-distance explicitly. But it is not that well worked-out, as covering only non-relativistic quantum mechanics; for all that Bell promoted it, no significant progress has been made in relativistic pilot-wave theory since his day – or for that matter since it was first proposed by de Broglie, in 1927 (challenge: provide a pilot-wave model of pair-creation in quantum electrodynamics). It is hard to see how this rejection of LOC really engages with Bell's concerns, or indeed Einstein's, when there is a clear alternative that respects them. Yet I cannot help but feel that this is the answer that Bell would have liked the least. (Einstein, I'm not so sure.)

**Acknowledgments**

My thanks to Chares Bédard, Dorje Brody, Jeremy Butterfield, Wayne Myrvold, Paul Tappenden, James Read, Cristi Stoica, Christopher Timpson, David Wallace, and especially to Harvey Brown, for setting me right in several places; errors that remain are my own. I am also grateful to Alyssa Ney, without whom this piece would not have been written.